\documentclass[conference,a4paper]{IEEEtran}

\usepackage[T1]{fontenc}
\usepackage[utf8]{inputenc}
\usepackage{times}
\usepackage{amsmath,amssymb}
\usepackage{graphicx}
\usepackage[table]{xcolor}
\usepackage{booktabs}
\usepackage{array}
\usepackage{tabularx}
\usepackage{multirow}
\usepackage{cite}
\usepackage{url}
\usepackage{tikz}
\usepackage{pgfplots}
\usepackage{stfloats}
\usepackage{float}
\usepackage{placeins}
\usepackage{balance}
\pgfplotsset{compat=1.18}
\usetikzlibrary{shapes.geometric,arrows.meta,positioning,fit,calc,backgrounds,patterns}

\title{AIRO: A Streaming-Based Machine Learning Framework for Threat Detection and Security in Industrial Internet of Things}

\author{
    \IEEEauthorblockN{
        Raima Deb\IEEEauthorrefmark{1},
        Pranav M Pawar\IEEEauthorrefmark{1},
       Mithun Mukherjee\IEEEauthorrefmark{1},
    Raja Muthalagu\IEEEauthorrefmark{1} and
    Mian Guo\IEEEauthorrefmark{2}
    }
    \IEEEauthorblockA{\IEEEauthorrefmark{1}Department of Computer Science, Birla Institute of Technology and Science, Pilani, Dubai Campus,\\
    Dubai International Academic City, Dubai, United Arab Emirates\\
    emails: \{p20230011, pranav, mithun, raja.m\}
    @dubai.bits-pilani.ac.in}
    \IEEEauthorblockA{\IEEEauthorrefmark{2}
    Guangdong Polytechnic Normal University, Guangzhou, China,
    Email: mian.guo123@gmail.com}
    
}

\begin{document}
\maketitle
\vspace{-1.0em}

\begin{abstract}
Modern Industrial Internet of Things (IIoT) implementations now connect edge gateways, cloud dashboards and business applications. This improves visibility and remote monitoring. However it simultaneously exposes industrial traffic to various cyber threats including scanning, web attacks and Structured Query Language (SQL) injection (SQLi), phishing-driven credential theft, and command-and-control (C2) activity. Although rule-based tools remain useful for detecting known signatures, their efficacy decreases with minimally altered attacks or when alerts require manual investigation.
This paper presents AIRO (Autonomous Intelligence-Driven Cybersecurity Risk Reduction System), an efficient framework for IIoT monitoring and response. AIRO combines Apache Kafka streaming, attack-specific feature extraction, machine learning (ML) inference, confidence-based decision logic and risk reduction measures such as Internet Protocol (IP) blocking, account suspension and traffic limiting.
We developed four specialized datasets for C2 beaconing, reconnaissance, SQLi, and phishing to evaluate AIRO. These datasets include noisy and overlapping characteristic distributions to minimize label leakage. The best-performing models, primarily Gradient Boosting and LightGBM, achieve F1-scores up to 0.9229 while ensuring low-latency from event input to mitigation.

\end{abstract}


\section{Introduction}

IIoT systems are common in manufacturing, energy, logistics, smart buildings and critical infrastructure. At a site, sensors and controllers do not work by themselves anymore. They share data with edge computers, dashboards, analytics platforms and remote maintenance tools. This is very helpful for automation. However it also makes industrial assets more vulnerable to threats that used to affect enterprise information technology networks~\cite{ref1}, \cite{refBotscan}.

The main motivation for this work is that we need a security mechanism that can monitor traffic, find behavior quickly and take action before a problem affects production. The usual way of detecting intrusions using signatures is helpful for known threats. However it is not very good, at stopping malware, slow scanning, special SQLi payloads, phishing campaigns, and C2 activity that are always changing and evolving. When alerts come in from sensors, gateways and user endpoints, manual response is also slow. We need something to protect these systems~\cite{ref4}, \cite{ref10}, \cite{ref5}.

This paper proposes AIRO, an autonomous intelligence-driven framework for IIoT threat
detection and mitigation. The work has four goals: to design a streaming security
pipeline, prepare attack-specific datasets, compare suitable ML
models, and connect detection output to practical response actions. The main
contributions are: (i) an IIoT-oriented architecture that combines Kafka streaming,
feature extraction, ML inference, and decision logic; (ii) four datasets for C2
beaconing~\cite{refBotscan}, \cite{ref11}, reconnaissance~\cite{ref1}, \cite{ref3}, SQLi~\cite{ref2}, \cite{ref10}, and phishing~\cite{ref5}, \cite{ref9}; (iii) a comparative
evaluation of classical and ensemble ML models; and (iv) a risk-scoring policy
that automates high-confidence responses while routing uncertain cases to analyst
review.

The rest of the paper is organized as follows. Section~II reviews related work and
summarizes the research gap. Section~III presents the proposed methodology in the
order suggested for the design: system architecture and its explanation, Kafka
pipeline, and dataset preparation with dataset description. Section~IV reports
results and discussion. Section~V concludes the paper.

\section{Related Work}

ML-based intrusion detection has been widely studied as an alternative review classical Intrusion Detection System (IDS) datasets and algorithms like Random Forest and Support Vector Machine (SVM). They identify common issues such as outdated traffic traces, class imbalance, and limited adaptability to new attacks. Ghourabi~\cite{ref2} demonstrated the effectiveness of hybrid LightGBM and Transformer models~\cite{ref6} for
healthcare fog security, noting their ability to capture complex patterns. However this study was domain-specific and focused solely on detection. Recent surveys and frameworks in the Internet of Vehicles (IoV) and Internet of Things (IoT)~\cite{ref3}, \cite{ref6}, \cite{ref8} confirm that ensemble learning and
feature selection improve classification performance in connected environments.

Cloud and network security research increasingly combines detection with partial
response. Mohammed \textit{et al.}~\cite{ref4} proposed an AI-powered incident
detection and response platform for cloud environments. Ghani
\textit{et al.}~\cite{ref10} used Isolation Forest and Random Forest to detect distributed denial-of-service (DDoS), malware, and backdoor attacks. Although valuable for operational
security, the response logic in these studies is typically limited and not structured around confidence-based autonomous mitigation. Kim \textit{et al.}~\cite{ref5} and
Malatji~\cite{ref9} emphasized the importance of human involvement and analyst trust for safety. However overreliance on manual review can reduce response speed in IIoT settings.

Recent research also addresses advanced anomaly correlation. Kumar
\textit{et al.}~\cite{ref7} integrated Isolation Forest, Autoencoders and convolutional long short-term memory (ConvLSTM)
models for robust anomaly detection. Zhu \textit{et al.}~\cite{ref11} introduced
LinTracer for tracking multi stage attack chains using graph-based correlation.
Ness \textit{et al.}~\cite{refNess} evaluated advanced ML models for network anomaly
detection. However their focus is offline classification rather than a closed-loop
real-time pipeline. BotScan~\cite{refBotscan} demonstrated the value of lightweight AI defense for IoT botnet threats and supports the need for deployable edge aware security mechanisms.

Existing works show that ML and
ensemble models are effective for intrusion detection. However many studies either depend on
older datasets or focus on one domain or stop at detection. A practical IIoT solution
requires streaming ingestion, attack-specific features, low-latency inference,
confidence-aware decisions, and automated mitigation. AIRO addresses this gap by
integrating detection and response in a single IIoT-oriented framework.

\section{Proposed Methodology}

\subsection{System Architecture and Explanation}

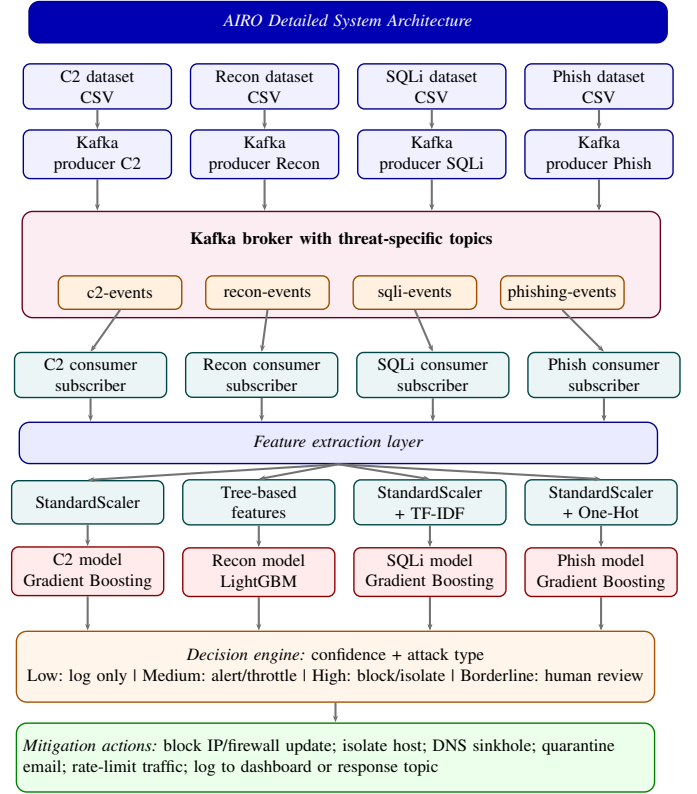
\begin{figure}[t]
\centering
\resizebox{\columnwidth}{!}{%
\begin{tikzpicture}[
  font=\tiny,
  stage/.style={rectangle, rounded corners=2pt, draw=blue!55!black, fill=blue!6,
                text width=1.38cm, align=center, minimum height=0.42cm, inner sep=1.8pt},
  wide/.style={rectangle, rounded corners=3pt, draw=blue!55!black, fill=blue!8,
               text width=6.35cm, align=center, minimum height=0.42cm, inner sep=2pt},
  broker/.style={rectangle, rounded corners=3pt, draw=purple!55!black, fill=purple!7,
                 text width=6.35cm, align=center, minimum height=1.08cm, inner sep=2pt},
  topic/.style={rectangle, rounded corners=2pt, draw=orange!60!black, fill=orange!12,
                text width=1.15cm, align=center, minimum height=0.34cm, inner sep=1.5pt},
  branch/.style={rectangle, rounded corners=2pt, draw=teal!55!black, fill=teal!8,
                 text width=1.42cm, align=center, minimum height=0.44cm, inner sep=1.8pt},
  model/.style={rectangle, rounded corners=2pt, draw=red!55!black, fill=red!8,
                text width=1.42cm, align=center, minimum height=0.44cm, inner sep=1.8pt},
  decision/.style={rectangle, rounded corners=3pt, draw=orange!65!black, fill=orange!8,
                   text width=6.35cm, align=center, minimum height=0.72cm, inner sep=2pt},
  mitigation/.style={rectangle, rounded corners=3pt, draw=green!50!black, fill=green!8,
                     text width=6.35cm, align=left, minimum height=0.70cm, inner sep=2pt},
  arr/.style={-{Stealth[length=2pt]}, semithick, black!55}
]
\node[wide, fill=blue!70!black, text=white] (title) {\textit{AIRO Detailed System Architecture}};

\node[stage, below=0.20cm of title, xshift=-2.55cm] (d1) {C2 dataset\\CSV};
\node[stage, right=0.18cm of d1] (d2) {Recon dataset\\CSV};
\node[stage, right=0.18cm of d2] (d3) {SQLi dataset\\CSV};
\node[stage, right=0.18cm of d3] (d4) {Phish dataset\\CSV};

\node[stage, below=0.20cm of d1] (p1) {Kafka\\producer C2};
\node[stage, below=0.20cm of d2] (p2) {Kafka\\producer Recon};
\node[stage, below=0.20cm of d3] (p3) {Kafka\\producer SQLi};
\node[stage, below=0.20cm of d4] (p4) {Kafka\\producer Phish};

\node[broker, below=0.32cm of p2, xshift=0.78cm] (broker) {};
\node[font=\tiny\bfseries, below=0.09cm of broker.north] (brokerlabel) {Kafka broker with threat-specific topics};
\node[topic, below=0.16cm of brokerlabel, xshift=-2.25cm] (t1) {c2-events};
\node[topic, right=0.23cm of t1] (t2) {recon-events};
\node[topic, right=0.23cm of t2] (t3) {sqli-events};
\node[topic, right=0.23cm of t3] (t4) {phishing-events};

\node[branch, below=0.34cm of broker, xshift=-2.55cm] (c1) {C2 consumer\\subscriber};
\node[branch, right=0.18cm of c1] (c2) {Recon consumer\\subscriber};
\node[branch, right=0.18cm of c2] (c3) {SQLi consumer\\subscriber};
\node[branch, right=0.18cm of c3] (c4) {Phish consumer\\subscriber};

\node[wide, below=0.24cm of c2, xshift=0.78cm] (featurelabel) {\textit{Feature extraction layer}};
\node[branch, below=0.16cm of featurelabel, xshift=-2.55cm] (f1) {StandardScaler};
\node[branch, right=0.18cm of f1] (f2) {Tree-based\\features};
\node[branch, right=0.18cm of f2] (f3) {StandardScaler\\+ TF-IDF};
\node[branch, right=0.18cm of f3] (f4) {StandardScaler\\+ One-Hot};

\node[model, below=0.22cm of f1] (m1) {C2 model\\Gradient Boosting};
\node[model, below=0.22cm of f2] (m2) {Recon model\\LightGBM};
\node[model, below=0.22cm of f3] (m3) {SQLi model\\Gradient Boosting};
\node[model, below=0.22cm of f4] (m4) {Phish model\\Gradient Boosting};

\node[decision, below=0.34cm of m2, xshift=0.78cm] (decision) {\textit{Decision engine:} confidence + attack type\\Low: log only \textbar{} Medium: alert/throttle \textbar{} High: block/isolate \textbar{} Borderline: human review};
\node[mitigation, below=0.20cm of decision] (mitigation) {\textit{Mitigation actions:} block IP/firewall update; isolate host; DNS sinkhole; quarantine email; rate-limit traffic; log to dashboard or response topic};

\foreach \from/\to in {d1/p1,d2/p2,d3/p3,d4/p4}{
  \draw[arr] (\from.south) -- (\to.north);
}
\foreach \from in {p1,p2,p3,p4}{
  \draw[arr] (\from.south) -- (\from.south |- broker.north);
}
\foreach \from/\to in {t1/c1,t2/c2,t3/c3,t4/c4}{
  \draw[arr] (\from.south) -- (\to.north);
}
\foreach \from in {c1,c2,c3,c4}{
  \draw[arr] (\from.south) -- (\from.south |- featurelabel.north);
}
\foreach \from/\to in {featurelabel/f1,featurelabel/f2,featurelabel/f3,featurelabel/f4,f1/m1,f2/m2,f3/m3,f4/m4}{
  \draw[arr] (\from.south) -- (\to.north);
}
\foreach \from in {m1,m2,m3,m4}{
  \draw[arr] (\from.south) -- (\from.south |- decision.north);
}
\draw[arr] (decision.south) -- (mitigation.north);
\end{tikzpicture}}
\caption{Detailed AIRO system architecture organized as a block diagram from datasets and Kafka streaming to model inference and mitigation.}
\label{fig:architecture}
\end{figure}

AIRO is designed as a layered IIoT security framework, as shown in
Fig.~\ref{fig:architecture}. Field devices, gateways, web applications, Domain Name System (DNS) logs,
authentication systems, and email security tools generate events continuously. These
events are normalized and sent to Kafka topics. During experimentation,
comma-separated values (CSV) files emulate live streams, and SQLi payload text is
represented using term frequency--inverse document frequency (TF-IDF) features~\cite{ref2}. The
inference layer extracts attack-specific features and applies the selected model.
The decision engine then maps model confidence to an operational response.

Fig.~\ref{fig:architecture} presents the detailed component-level architecture
of AIRO. \textit{The data-stream layer} contains the four attack data streams used in the
prototype: C2 beaconing, reconnaissance, SQLi, and phishing. Each stream is
represented as a CSV dataset during experimentation, but the same position can later
be occupied by live IIoT logs, DNS records, web-application events, authentication
traces, or email-security alerts.

\textit{The data-ingestion layer} uses four independent Kafka producers. Each producer reads
from its attack-specific source, serializes every record as a JavaScript Object Notation (JSON) message, and
publishes it to the matching Kafka topic. In the prototype, the producer rate is
kept at about one event every 0.3 seconds per stream so that the pipeline behaves
like a live monitoring system instead of a one-time batch script. The JSON record
keeps raw feature columns together with metadata such as source IP, destination IP,
host, and timestamp.

\textit{The Kafka broker layer} maintains four isolated topics: \texttt{c2-events},
\texttt{recon-events}, \texttt{sqli-events}, and \texttt{phishing-events}. This
separation prevents a high-volume stream, such as phishing traffic, from delaying a
low-latency stream such as C2 detection. It also allows each topic to be replayed
from Kafka logs if a consumer fails or if a model has to be tested again on the same
sequence of events.

\textit{The consumer and feature-extraction layers} convert incoming topic messages into
model-ready inputs. C2 records use StandardScaler for timing and beaconing features,
reconnaissance records use tree-based network features, SQLi records use
StandardScaler with TF-IDF payload features, and phishing records use StandardScaler
with one-hot encoded Uniform Resource Locator (URL) or login attributes. These preprocessing steps produce the
fixed-length vectors expected by the trained classifiers.

\textit{The ML inference layer} contains four dedicated inference threads. Each consumer loads the
corresponding pre-trained \texttt{.pkl} model and calls probability inference for
each event. The output is a confidence score between 0 and 1. \textit{The decision and mitigation layer} then maps this score and the attack type to an action: scores below 0.50 are logged,
0.50--0.80 produces an alert or rate limit, scores of 0.80 and above trigger stronger
mitigation such as blocking, isolation, sinkholing, or quarantine, and the borderline
band of 0.45--0.55 is escalated for human review. This layer keeps the system
practical for IIoT use because risky actions are linked to both confidence and
operational impact.

\subsection{Kafka Pipeline}

\begin{figure}[t]
\centering
\resizebox{1.05\columnwidth}{!}{%
\begin{tikzpicture}[
  font=\scriptsize,
  box/.style={rectangle, rounded corners=3pt, draw=blue!45!black, fill=blue!7,
              text width=1.54cm, align=center, minimum height=0.58cm, inner sep=2pt},
  topic/.style={rectangle, rounded corners=2pt, draw=orange!55!black, fill=orange!9,
                text width=1.40cm, align=center, minimum height=0.46cm, inner sep=2pt},
  action/.style={rectangle, rounded corners=3pt, draw=green!45!black, fill=green!8,
                 text width=1.56cm, align=center, minimum height=0.58cm, inner sep=2pt},
  arr/.style={-{Stealth[length=3pt]}, semithick, black!55}
]
\node[box] (sources) {IIoT logs\\and events};
\node[box, right=0.18cm of sources] (producer) {Kafka\\producers};
\node[topic, right=0.20cm of producer, yshift=0.62cm] (c2) {c2-events};
\node[topic, below=0.12cm of c2] (recon) {recon-events};
\node[topic, below=0.12cm of recon] (sqli) {sqli-events};
\node[topic, below=0.12cm of sqli] (phish) {phishing-events};
\node[box, right=0.22cm of recon, yshift=-0.31cm] (consumer) {Consumer group\\feature extraction};
\node[box, right=0.18cm of consumer] (model) {ML model\\inference};
\node[topic, right=0.18cm of model] (response) {Response\\topic};
\node[action, right=0.18cm of response] (decision) {Decision and\\mitigation};
\draw[arr] (sources)--(producer);
\draw[arr] (producer.east)--(c2.west);
\draw[arr] (producer.east)--(recon.west);
\draw[arr] (producer.east)--(sqli.west);
\draw[arr] (producer.east)--(phish.west);
\draw[arr] (c2.east)--(consumer.west);
\draw[arr] (recon.east)--(consumer.west);
\draw[arr] (sqli.east)--(consumer.west);
\draw[arr] (phish.east)--(consumer.west);
\draw[arr] (consumer)--(model);
\draw[arr] (model)--(response);
\draw[arr] (response)--(decision);
\end{tikzpicture}}
\caption{Kafka pipeline connecting event ingestion, model inference, and response.}
\label{fig:kafkapipeline}
\end{figure}
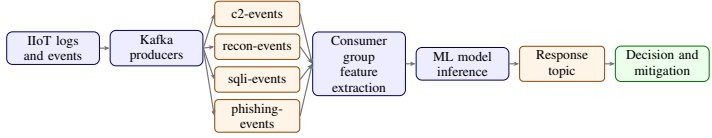

Kafka is used as the real-time backbone that connects data generation, feature
extraction, model inference, and mitigation. Four producers feed the four dedicated
topics shown in Fig.~\ref{fig:kafkapipeline}. Each topic is consumed by a matching
consumer thread, which deserializes the JSON record, applies the required
preprocessing step, loads the relevant \texttt{.pkl} model, and produces a confidence
score. The broker also provides simple fault tolerance because retained topic logs
allow replay when a consumer restarts. In the prototype, the end-to-end delay from
event arrival to decision is approximately 0.5 seconds.

\begin{table}[t]
\centering
\caption{Kafka event schema used by the prototype.}
\label{tab:schema}
\footnotesize
\begin{tabular}{p{2.0cm}p{4.9cm}}
\toprule
Field & Description \\
\midrule
\texttt{event\_id} & Unique identifier for traceability and duplicate handling. \\
\texttt{timestamp} & Time at which the gateway or connector observed the event. \\
\texttt{asset\_type} & programmable logic controller (PLC), sensor gateway, web server, user endpoint, or cloud service. \\
\texttt{src}, \texttt{dst} & Source and destination address, domain, or service identifier. \\
\texttt{host} & Device, endpoint, or service that generated the event. \\
\texttt{features} & Extracted numerical or encoded features used by the model. \\
\texttt{score} & Probability returned by the selected classifier. \\
\texttt{action} & Response selected by the decision engine. \\
\bottomrule
\end{tabular}
\end{table}

The schema in Table~\ref{tab:schema} keeps the streaming record small enough for
high-volume traffic while retaining the fields needed for auditing. Raw packet
payloads do not need to be stored permanently in the Kafka topic; instead, compact
features and references are retained. This reduces storage cost and limits exposure
of sensitive operational data. If a serious incident is confirmed, the event
identifier can be used to retrieve deeper logs from the original sensor or gateway.

Fault tolerance is handled through Kafka offsets. If an inference consumer fails,
it can restart from the last committed offset and continue processing without losing
events. This behaviour is useful in IIoT sites where connectivity between edge
locations and central monitoring may be unstable. The system can also prioritize
local enforcement for high-risk events, allowing a gateway to take temporary action
even if the central dashboard is delayed.

Mitigation is intentionally confidence-driven. For example, a repeated scan from an
unknown source can trigger temporary IP blocking, a phishing login attempt can lock
or challenge the account, a SQLi attempt can generate a web-application firewall (WAF) rule, and suspected C2 traffic can be redirected to a DNS sinkhole. Events
near the decision boundary are escalated to a human analyst to reduce the risk of
unnecessary disruption in industrial operations.

\begin{table}[t]
\centering
\caption{Decision policy used by the AIRO response engine.}
\label{tab:policy}
\footnotesize
\begin{tabular}{p{1.5cm}p{1.55cm}p{3.55cm}}
\toprule
Score range & Risk level & Action in IIoT deployment \\
\midrule
$p < 0.45$ & Low & Store event, update dashboard, no automated disruption. \\
$0.45 \le p \le 0.55$ & Borderline & Notify analyst and request confirmation before enforcement. \\
$0.55 < p < 0.80$ & Medium & Raise alert, increase logging, throttle repeated sources. \\
$p \ge 0.80$ & High & Apply attack-specific block, isolation, sinkhole, or quarantine. \\
\bottomrule
\end{tabular}
\end{table}

The response policy in Table~\ref{tab:policy} is conservative because industrial
networks value availability as much as confidentiality. A high-confidence SQLi event can be blocked at the web gateway without stopping production
traffic, whereas isolating an engineering workstation requires stronger evidence.
For this reason, the same probability score is interpreted together with asset
criticality, event repetition, and source reputation. The framework therefore avoids
a single ``detect and block'' rule and instead selects a response that matches the
operational risk.

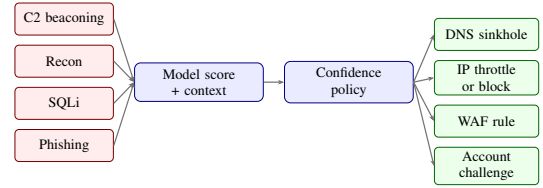
\begin{figure}[t]
\centering
\scalebox{0.77}{%
\begin{tikzpicture}[
  font=\scriptsize,
  threat/.style={rectangle, rounded corners=2pt, draw=red!45!black, fill=red!7,
                 text width=1.55cm, align=center, minimum height=0.55cm, inner sep=2pt},
  engine/.style={rectangle, rounded corners=3pt, draw=blue!50!black, fill=blue!8,
                 text width=2.0cm, align=center, minimum height=0.65cm, inner sep=3pt},
  action/.style={rectangle, rounded corners=2pt, draw=green!40!black, fill=green!8,
                 text width=1.65cm, align=center, minimum height=0.55cm, inner sep=2pt},
  arr/.style={-{Stealth[length=3pt]}, semithick, black!55}
]
\node[threat] (c2) {C2 beaconing};
\node[threat, below=0.16cm of c2] (recon) {Recon};
\node[threat, below=0.16cm of recon] (sqli) {SQLi};
\node[threat, below=0.16cm of sqli] (phish) {Phishing};
\node[engine, right=0.35cm of recon, yshift=-0.36cm] (score) {Model score\\+ context};
\node[engine, right=0.35cm of score] (policy) {Confidence\\policy};
\node[action, right=0.35cm of policy, yshift=0.82cm] (sink) {DNS sinkhole};
\node[action, below=0.16cm of sink] (block) {IP throttle\\or block};
\node[action, below=0.16cm of block] (waf) {WAF rule};
\node[action, below=0.16cm of waf] (lock) {Account\\challenge};
\draw[arr] (c2.east)--(score.west);
\draw[arr] (recon.east)--(score.west);
\draw[arr] (sqli.east)--(score.west);
\draw[arr] (phish.east)--(score.west);
\draw[arr] (score)--(policy);
\draw[arr] (policy.east)--(sink.west);
\draw[arr] (policy.east)--(block.west);
\draw[arr] (policy.east)--(waf.west);
\draw[arr] (policy.east)--(lock.west);
\end{tikzpicture}}
\caption{Threat-specific mapping from detection score to mitigation control.}
\label{fig:mitigationmap}
\end{figure}

Fig.~\ref{fig:mitigationmap} keeps the response path explicit. It is included
because several reference papers use architecture and process diagrams to explain
how detection output becomes an operational security action. In AIRO,
the same idea is adapted for IIoT by separating the threat-specific detector from
the enforcement point. This prevents the model from producing only an abstract
label and makes the final action easier to audit.

For example, consider an IIoT gateway that normally communicates with a small set of
vendor cloud endpoints. If the gateway starts sending periodic DNS queries to a new
low-reputation domain and the timing resembles a beacon, the C2 model assigns a high
score. The decision engine can then add the domain to a sinkhole list while keeping
normal plant traffic active. In contrast, if an engineer's workstation performs a
short port scan during an approved maintenance window, the event may remain in the
borderline region and be routed to review instead of being blocked immediately.
This distinction is essential because industrial security controls must avoid
unnecessary downtime.

From an implementation point of view, every event has some details. These details include a timestamp, source, destination, what kind of asset is involved, some features that were extracted, what the model output is, and the status of the action taken. This information is sent back to a response topic. This way, dashboards, audit reports, and people handling incidents can look at the whole sequence of events. Keeping a decision history within the pipeline is helpful. It helps adjust the thresholds after deploying the system. False positives and analyst overrides can be reviewed without manually collecting logs from multiple tools.

\subsection{Dataset Preparation and Description}

Custom datasets were prepared for four IIoT
threat categories because public datasets often contain out-of-date attacks, duplicate entries, or
features that leak the label, . The generation process used realistic ranges, noise, overlap
between benign and malicious classes, and adversarial variants. Each dataset was divided into training and test sets, and categorical or textual attributes were
encoded before model training. Table~\ref{tab:datasets} summarizes the size, class distribution, and feature type for each dataset used in the prototype.

\begin{table}[t]
\centering
\caption{Dataset overview used in the prototype.}
\label{tab:datasets}
\footnotesize
\begin{tabular}{p{1.35cm}p{0.62cm}p{0.62cm}p{0.62cm}p{2.75cm}}
\toprule
Dataset & Total & Mal. & Norm. & Features \\
\midrule
C2 beaconing & 5,000 & 1,264 & 3,736 & 10 numerical \\
Recon. & 6,000 & 1,898 & 4,102 & 12 numerical \\
SQLi & 5,000 & 1,497 & 3,503 & 5 num. + TF-IDF \\
Phishing & 10,000 & 5,092 & 4,908 & 9 num. + 9 cat. \\
\bottomrule
\end{tabular}
\end{table}

\textit{C2 beaconing dataset:} This dataset is made up of network behavior that is related to command and control callbacks. Every sample is an observation period for each host. Normal samples include telemetry, update requests, and cloud synchronization, while the malicious samples involve periodic callbacks, low-jitter inter-arrival time, abnormal DNS bursts, and low-reputation destinations. There are ten numerical features that describe timing, size, DNS entropy, jitter, and destination reputation, as shown in Table~\ref{tab:c2features}.

\textit{Reconnaissance dataset:} This dataset is about port scanning and asset discovery on IIoT gateway and services. In the benign class, there will be maintenance scans and routine service checks to avoid the situation where each scan is regarded as an attack. The malicious class has high unique port numbers, increased failure connections, more port entropy, odd scan rate, and stronger sequential scan. Twelve numerical variables are considered for this class, which are provided in Table~\ref{tab:reconfeatures}.

\textit{SQLi dataset:} The data set consists of web requests labeled as either normal or SQLi attack attempts. The normal requests have the following features: form submission and parameterized requests; while the malicious requests include tautologies, union queries, comments, encoded requests, errors-based attacks, and obfuscation attacks. Five numerical metadata features are accompanied by the TF-IDF features calculated based on the query, including special characters, because they help detect attacks. The numerical and textual SQLi
features are provided in Table~\ref{tab:sqlifeatures}.

\textit{Phishing dataset:} The dataset consists of interactions for phishing
activities of users and emails. The features considered include URL, sender and domain attributes, login context, and simple network characteristics. Malicious instances include suspect tokens, URL redirection, low domain reputation, and abnormal login behavior. The legitimate instances consist of cloud service and enterprise URLs. These are used in order to teach the classifier something more than URL length and token number. It is summarized in Table~\ref{tab:phishingfeatures}.

\begin{table}[t]
\centering
\caption{C2 beaconing feature descriptions.}
\label{tab:c2features}
\tiny
\begin{tabular}{p{2.75cm}p{3.95cm}}
\toprule
Feature & Description \\
\midrule
\texttt{num\_nodes} & Nodes in the network window. \\
\texttt{num\_malicious\_nodes} & Malicious nodes in the scenario. \\
\texttt{heterogeneity} & Diversity of node or device types. \\
\texttt{interval\_mean/std} & Mean and variation of beacon intervals. \\
\texttt{packet\_size} & Average packet size. \\
\texttt{dns\_requests} & DNS query count and burstiness. \\
\texttt{failed\_connections} & Failed outbound attempts. \\
\texttt{connection\_duration} & Average connection duration. \\
\bottomrule
\end{tabular}
\end{table}

\begin{table}[t]
\centering
\caption{Reconnaissance feature descriptions.}
\label{tab:reconfeatures}
\tiny
\begin{tabular}{p{2.75cm}p{3.95cm}}
\toprule
Feature & Description \\
\midrule
\texttt{connection\_attempts} & Total connection attempts. \\
\texttt{unique\_ports} & Distinct ports contacted. \\
\texttt{failed/success\_connections} & Failed and successful connections. \\
\texttt{scan\_rate} & Attempts per second. \\
\texttt{avg\_connection\_duration} & Average connection duration. \\
\texttt{packet\_rate}, \texttt{bytes\_sent} & Packet and byte volume. \\
\texttt{port\_entropy} & Randomness of port selection. \\
\texttt{sequential\_scan\_score} & Likelihood of ordered scanning. \\
\texttt{burstiness}, \texttt{interval\_variance} & Timing burst and interval variation. \\
\bottomrule
\end{tabular}
\end{table}

\begin{table}[t]
\centering
\caption{SQLi feature descriptions.}
\label{tab:sqlifeatures}
\scriptsize
\begin{tabular}{p{2.15cm}p{4.55cm}}
\toprule
Feature & Description \\
\midrule
\texttt{request\_rate} & Number of queries per second. \\
\texttt{failed\_queries} & Number of failed query attempts. \\
\texttt{response\_time} & Average query response time. \\
\texttt{payload\_length} & Query-string length. \\
\texttt{encoding\_flag} & URL or hex encoding indicator. \\
\texttt{query} & Raw SQL text transformed using TF-IDF, with special characters preserved. \\
\bottomrule
\end{tabular}
\end{table}

\begin{table}[t]
\centering
\caption{Phishing feature descriptions.}
\label{tab:phishingfeatures}
\scriptsize
\begin{tabular}{p{2.15cm}p{4.55cm}}
\toprule
Feature group & Description \\
\midrule
URL/domain & URL entropy, domain age, redirects, shortened links, and suspicious tokens are analyzed. \\
Sender & Sender reputation, mismatch flags, and email metadata are evaluated. \\
Login & Login location, frequency, and abnormal access behavior are monitored. \\
Categorical & Encoded sender, domain, and action categories are identified. \\
Network fields & Connection and request indicators near the email event are examined. \\
\bottomrule
\end{tabular}
\end{table}

Preprocessing included handling missing values, scaling numerical features,
encoding categorical variables and extracting text-derived features from URLs and payload strings. Three preparation schemes were used to accommodate the diverse data types of attack sources. First was StandardScaler to keep numerical timing and traffic features on a comparable scale. Second was TF-IDF to convert SQLi payload text into model-ready token weights. Last was using one-hot encoding to represent phishing categorical fields without imposing a false numeric order.
The dataset construction principles in Table~\ref{tab:principles}
summarize how overlap, noise, attack variants and benign look-alikes were retained
to avoid unrealistic separability. Train-test splitting was performed before final
evaluation to avoid information leakage. The datasets were not made perfectly
separable. Borderline records were retained so that model performance reflects realistic operational conditions.

\begin{table}[t]
\centering
\caption{Dataset construction principles.}
\label{tab:principles}
\footnotesize
\begin{tabular}{p{2.0cm}p{5.0cm}}
\toprule
Principle & Purpose \\
\midrule
Class overlap & Protects against the learning of impractical threshold policies. \\
Noise injection & Symbolizes packet losses, logging discontinuities, encoded strings, and measurement errors. \\
Attack variants & Comprises of low frequency scans, jittered C2, SQLi code, and short URLs. \\
Benign look-alikes & Includes software updates, maintenance scans, admin queries, and valid login URLs. \\
Before tuning split & Avoids accidental data leakage from training to evaluation sets. \\
\bottomrule
\end{tabular}
\end{table}

The records were reviewed from a security 
perspective rather than solely a statistical one. The
malicious class was not defined only by a fixed period for C2. It
also encompassed subtle timing changes common in real
malware. Benign maintenance windows
were included for reconnaissance because industrial teams regularly scan assets for inventory and compliance. Payloads
were encoded and integrated with normal parameters for SQLi whch prevented a
model from relying on a single symbol.
Legitimate cloud-service links were incorporated for phishing as many
false positives occur when URL-based detection automatically flags long
or redirected links as malicious.

This preparation is crucial for IIoT because public 
enterprise datasets rarely reflect the complex blend of operational
traffic, maintenance behavior and user-driven web activity
found in industrial environments. While these datasets do not replace live
capture, they offer a controlled setting for 
comparing models before deploying them to real devices.

Feature importance was also considered while interpreting the trained models. Timing consistency, DNS entropy, and destination reputation were among the most useful indicators for C2 dataset. Number of unique ports and failed connection ratio proved useful than just total packet count for reconnaissance. In SQLi, keyword frequency and encoded character patterns helped distinguish malicious
requests from normal parameters. No single feature was reliable enough for phishing which explains the lower F1-score and supports the need for richer text and domain
intelligence in future versions.

\section{Results and Discussion}

Logistic Regression, Decision Tree, Random Forest, Gradient Boosting, XGBoost and LightGBM were evaluated.
The evaluation used accuracy, precision, recall and F1-score.
The final selection focused on F1-score and stable test performance.
This is because we need both detection and controlled false positives. Table~\ref{tab:results}
summarizes the chosen model for each threat category.

The evaluation involved splitting the data preprocessing for each type of attack and practical model settings. Then the selected classifier was connected to its Kafka consumer.
The experimental setup and hyperparameters are listed in 
Table~\ref{tab:modelprep}.
Implementation files are available at
\url{https://github.com/pranav21684/Raima-Sem-2-2025-26}.

\begin{table}[t]
\centering
\caption{Experimental setup and hyperparameters.}
\label{tab:modelprep}
\tiny
\setlength{\tabcolsep}{1pt}
\begin{tabular}{p{0.75cm}p{1.05cm}p{1.55cm}p{1.20cm}p{1.30cm}}
\toprule
\shortstack{Threat} & \shortstack{Train/Val/\\Test split} & \shortstack{Pre-\\processing} & \shortstack{Model} & \shortstack{Hyper-\\parameters} \\
\midrule
C2 & 64/16/20 & StandardScaler numerical features & Gradient Boosting & 100 estimates, lr 0.1 \\
Recon. & 80/0/20 & Tree features, class weight 1:2 & LightGBM & 100 estimates, lr 0.1, 31 leaves \\
SQLi & 60/20/20 & StandardScaler, TF-IDF & Gradient Boosting & 100 estimates, lr 0.1 \\
Phishing & 70/15/15 & StandardScaler, one-hot encoding & Gradient Boosting & 100 estimates, lr 0.1 \\
\bottomrule
\end{tabular}
\end{table}

\begin{table}[t]
\centering
\caption{Best-performing models for AIRO datasets.}
\label{tab:results}
\footnotesize
\begin{tabular}{p{1.7cm}p{2.2cm}cccc}
\toprule
Threat & Selected model & Acc. & Prec. & Rec. & F1 \\
\midrule
C2 beaconing & Gradient Boosting & 0.923 & 0.914 & 0.932 & 0.923 \\
Reconnaissance & LightGBM & 0.914 & 0.909 & 0.918 & 0.913 \\
SQLi & Gradient Boosting & 0.908 & 0.897 & 0.916 & 0.906 \\
Phishing & Gradient Boosting & 0.671 & 0.660 & 0.641 & 0.650 \\
\bottomrule
\end{tabular}
\end{table}

The results show that ensemble methods work better than a linear model for these datasets. This makes sense because attack labels depend on combinations of features, not one threshold. The Receiver Operating Characteristic (ROC) curves show the area under the curve (AUC) values. SQLi is used in the plot labels. C2 and reconnaissance detection gave the clearest results.
This is because timing regularity, destination reputation, port fan-out and failed-connection ratios are behavioral signals. SQLi was also detected reliably.
However encoded payloads and normal requests, with characters caused some overlap for phishing category. This is because benign and malicious URLs can look similar. For this reason phishing is used as a warning component not as a stand-alone blocking mechanism.

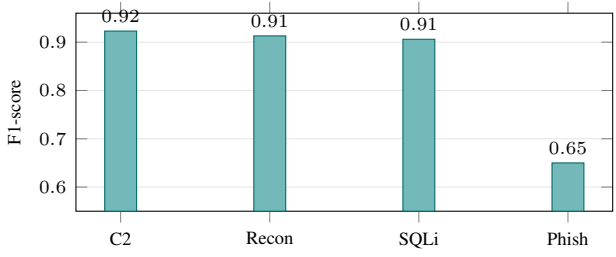
\begin{figure}[t]
\centering
\begin{tikzpicture}
\begin{axis}[
  ybar,
  width=0.98\columnwidth,
  height=4.2cm,
  ymin=0.55,ymax=0.96,
  ylabel={F1-score},
  symbolic x coords={C2,Recon,SQLi,Phish},
  xtick=data,
  nodes near coords,
  nodes near coords style={font=\scriptsize},
  bar width=12pt,
  tick label style={font=\scriptsize},
  label style={font=\scriptsize},
  ymajorgrids=true,
  grid style={black!10}
]
\addplot[fill=teal!55,draw=teal!80!black] coordinates
  {(C2,0.923) (Recon,0.913) (SQLi,0.906) (Phish,0.650)};
\end{axis}
\end{tikzpicture}
\caption{F1-score of chosen models for each threat category.}
\label{fig:f1summary}
\end{figure}

\begin{figure}[t]
\centering
\begin{tikzpicture}
\begin{axis}[
  ybar,
  width=0.98\columnwidth,
  height=4.2cm,
  ymin=0.55,ymax=0.96,
  ylabel={Precision},
  symbolic x coords={C2,Recon,SQLi,Phish},
  xtick=data,
  nodes near coords,
  nodes near coords style={font=\scriptsize},
  bar width=12pt,
  tick label style={font=\scriptsize},
  label style={font=\scriptsize},
  ymajorgrids=true,
  grid style={black!10}
]
\addplot[fill=orange!55,draw=orange!80!black] coordinates
  {(C2,0.914) (Recon,0.909) (SQLi,0.897) (Phish,0.660)};
\end{axis}
\end{tikzpicture}
\caption{Precision of selected models from each threat category.}
\label{fig:precisionsummary}
\end{figure}
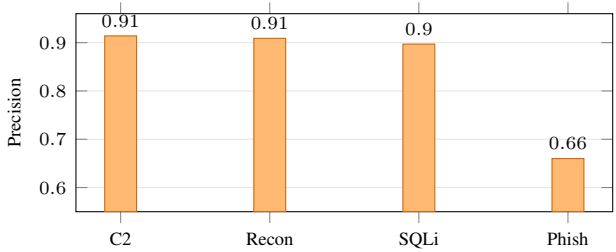

\begin{figure}[t]
\centering
\begin{tikzpicture}
\begin{axis}[
  width=0.98\columnwidth,
  height=4.8cm,
  xmin=0,xmax=1,
  ymin=0,ymax=1,
  xlabel={False positive rate},
  ylabel={True positive rate},
  tick label style={font=\scriptsize},
  label style={font=\scriptsize},
  legend style={font=\tiny, at={(0.98,0.03)}, anchor=south east, draw=none, fill=white},
  grid=both,
  grid style={black!10}
]
\addplot[black!45,dashed,semithick] coordinates {(0,0) (1,1)};
\addlegendentry{Baseline}
\addplot[teal!80!black,thick,mark=none] coordinates
  {(0,0) (0.03,0.42) (0.08,0.71) (0.16,0.88) (0.28,0.95) (1,1)};
\addlegendentry{C2, AUC=0.95}
\addplot[blue!75!black,thick,mark=none] coordinates
  {(0,0) (0.04,0.38) (0.10,0.69) (0.19,0.86) (0.32,0.94) (1,1)};
\addlegendentry{Recon, AUC=0.94}
\addplot[red!75!black,thick,mark=none] coordinates
  {(0,0) (0.05,0.34) (0.12,0.65) (0.23,0.84) (0.38,0.92) (1,1)};
\addlegendentry{SQLi, AUC=0.92}
\addplot[purple!75!black,thick,mark=none] coordinates
  {(0,0) (0.10,0.22) (0.25,0.47) (0.45,0.66) (0.68,0.82) (1,1)};
\addlegendentry{Phish, AUC=0.71}
\end{axis}
\end{tikzpicture}
\caption{ROC curves for the selected AIRO classifiers.}
\label{fig:rocsummary}
\end{figure}
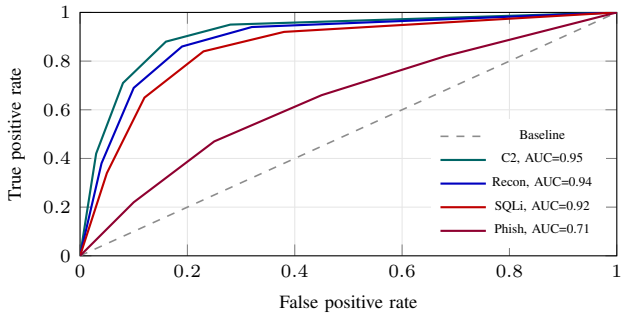

Figs.~\ref{fig:f1summary} and~\ref{fig:precisionsummary} shows that we get the results when the attack has stable patterns of behavior. The way C2 communicates leaves patterns in timing and destination. Reconnaissance creates patterns in connections. SQLi changes the structure of web requests. The ROC curves in Fig.~\ref{fig:rocsummary} support the conclusion:  the C2, reconnaissance and SQLi classifiers are close to the upper-left region
but phishing has a weaker curve because good and bad URLs can look very similar. So we should treat phishing detection as a step with email authentication, sandboxing and user verification rather than the only way to control it.

When we deploy this two things are very important not just how accurate it is. First we have to think about latency: if an alert comes after the malware has already done something it gives the attacker time to do more damage. Second we have to think about reversibility: it is safer to block or quarantine something temporarily. Have an analyst approve it rather than making permanent changes that could affect something that is being used for real work. So AIRO uses short response times and actions that can be tracked
so that operators can undo something if it affects a task.

Table~\ref{tab:relatedcompare} compares our work to recent works. The table shows the metric from each work and whether it can be used in real-time and can respond automatically.
In this table, Grey Wolf Optimization (GWO), convolutional neural network (CNN),
and Internet of Medical Things (IoMT) are written as acronyms so the cells are easy to read.

\begin{table}[t]
\centering
\caption{Comparative analysis with related works.}
\label{tab:relatedcompare}
\tiny
\setlength{\tabcolsep}{1.5pt}
\begin{tabular}{>{\raggedright\arraybackslash}p{0.9cm}>{\raggedright\arraybackslash}p{1.75cm}>{\centering\arraybackslash}p{0.85cm}>{\centering\arraybackslash}p{0.75cm}>{\centering\arraybackslash}p{0.65cm}>{\centering\arraybackslash}p{0.75cm}}
\toprule
Work & Method/domain & Accuracy & F1-score & Stream & Response \\
\midrule
Choukhairi \textit{et al.}~\cite{ref8} & Butterfly opt. ensemble; IoT/IDS & 99.795\% & 0.9979 & No & No \\
Abid \textit{et al.}~\cite{ref6} & GWO + CNN-LightGBM; IoT/IoMT & 95.24\% & 0.9951 & No & No \\
Ness \textit{et al.}~\cite{refNess} & ML anomaly detection; network & 85\% & 0.85 & No & No \\
\rowcolor{blue!10} AIRO & Kafka + attack-specific ML; IIoT & 95.3\% & 0.9229 & Yes & Yes \\
\bottomrule
\end{tabular}
\end{table}

The attack categories are separate because SQLi depends on request
structure. C2 depends on timing and destination behavior, and each threat maps to a different control point. For deployment AIRO can start in monitoring mode then
enable alerting, reversible throttling and controlled automation after analysts
review local false positives.

\section{Conclusion}

This paper introduced AIRO which is a real time IIoT cybersecurity framework that combines threat detection with autonomous mitigation. Kafka is used for streaming ingestion which helps with attack specific feature extraction, ensemble ML models and confidence based decision making. Four datasets were prepared for C2 beaconing, reconnaissance, SQLi, and phishing attacks. The models performed well in all but one category with phishing being detected as the area requiring further work. In conclusion, AIRO showcases that lightweight ML and a streaming framework could lead to more rapid, secure and practical security responses in IIoT environments.

The main limitations of AIRO include low F1-score in phishing attack
detection and additional effort required for maintaining different models. Future work will involve testing the pipeline on IIoT traffic, introducing transformer based features for better phishing detection and model distillation for implementation at edge gateways.


\balance
\scriptsize
\setlength{\itemsep}{0pt}
\setlength{\parsep}{0pt}

\end{document}